%% file: harmonic_counterexample.tex
\newcommand{\documentversion}{12pt}
\documentclass[\documentversion]{elsarticle}

\makeatletter
\def\ps@pprintTitle{%
   \let\@oddhead\@empty
   \let\@evenhead\@empty
   \let\@oddfoot\@empty
   \let\@evenfoot\@oddfoot
}
\makeatother

\journal{}

\input{my_package_list.tex}
\newboolean{isELS}
\setboolean{isELS}{true}
\newboolean{isSupplement}
\setboolean{isSupplement}{false}
\newboolean{isReview}
\setboolean{isReview}{false}
\input{new_commands.tex}

\begin{document}

\begin{frontmatter}



\title{Counterexamples to Proofs for Volumetric Parameterization of Topological Sweeps}


\corref{cor1}
\author[address1]{Caleb Goates}
\ead{calebgoates@gmail.com}

\author[address2]{Kendrick M. Shepherd}
\ead{kendrick_shepherd@byu.edu}

\cortext[cor1]{Corresponding author}
\address[address1]{Department of Civil and Construction Engineering, Brigham Young University, USA}

\begin{abstract}
	\input{sections/Abstract.tex}

\end{abstract}

%
%

\end{frontmatter}


\input{sections/PaperBody.tex}



\section*{Acknowledgements}
K. Shepherd and C. Goates were supported by National Science Foundation under Grant No. 2245491.  Any opinion, findings, and conclusions or recommendations expressed in this material are those of the authors and do not necessarily reflect the views of the National Science Foundation.

\bibliographystyle{modifiedplain}
\bibliography{Combined_Bibliography_master}


\end{document}

%% file: my_package_list.tex
\usepackage[dvipsnames]{xcolor}

\usepackage{amsfonts}
\usepackage{amssymb}
\usepackage{amsmath}
\usepackage{amsthm}
\usepackage{amscd}
\usepackage{graphicx}
\usepackage{color}
\usepackage{enumerate}
\usepackage{float}
\usepackage{subcaption}
\usepackage[titletoc,title]{appendix}
\usepackage{ragged2e}
\usepackage{hyperref}
\usepackage[font={small}]{caption}
\hypersetup{
  colorlinks   = true, 
  urlcolor     = blue, 
  linkcolor    = black, 
  citecolor   = black 
}
\usepackage[ruled,noend]{algorithm2e}
\usepackage{todonotes}
\usepackage{tikz,tikz-cd}
\usetikzlibrary{shapes,arrows}
\usetikzlibrary{positioning}
\usepackage{ifthen}
\usepackage{xstring}

\usepackage{algpseudocode}
\algdef{SE}
[OBJECT]
{Obj}
{EndObj}
[1]
{\textbf{object} \textsc{#1}}
{\textbf{end object}}

\usepackage{ulem}

\usepackage{rotating}

\usepackage{wrapfig}

\usepackage{tikz}
\usetikzlibrary{patterns}
\usetikzlibrary{math}
\usetikzlibrary{shapes}

%% file: new_commands.tex
\graphicspath{{./}{figures/}{Figures/}{../Actual/Figures/}{../Actual/}}

\ifthenelse{\boolean{isReview}}{
	
	\newcommand{\footerfontsize}{\large}
	}
	{
	
	\newcommand{\footerfontsize}{\footnotesize}
	}

\definecolor{forestgreen}{cmyk}{0.76,0,0.76,0.45}

\definecolor{lightgray}{cmyk}{0.2,0.2,0.2,0.2}

\definecolor{burntorange}{rgb}{0.74902,0.341176,0}

\definecolor{myBlue}{rgb} {0,    0.4470,    0.7410}

\definecolor{armygreen}{rgb}{0.29, 0.33, 0.13}

\ifthenelse{\boolean{isELS}}{

	\theoremstyle{remark}
	
	}
	{}

\makeatletter
\newcommand{\tpitchfork}{%
  \vbox{
    \baselineskip\z@skip
    \lineskip-.52ex
    \lineskiplimit\maxdimen
    \m@th
    \ialign{##\crcr\hidewidth\smash{$-$}\hidewidth\crcr$\pitchfork$\crcr}
  }%
}
\makeatother

\makeatletter
\newcounter{datastruct}

\makeatother

\newcommand{\node}{node}

\usepackage{adjustbox}
\usepackage{array}

\newcolumntype{R}[2]{%
    >{\adjustbox{angle=#1,lap=\width-(#2)}\bgroup}%
    l%
    <{\egroup}%
}



%% file: sections/Abstract.tex
{\footerfontsize
Harmonic maps are important in generating parameterizations for various domains, particularly in two and three dimensions.
General extensions of two-dimensional harmonic parameterizations for volumetric parameterizations are known to fail in a variety of contexts, though more specialized volumetric parameterizations have been proposed.
This work provides and contextualizes a counterexample to various proposed proofs that employ harmonic maps to sweep a parameterization from a base surface, $\Gamma_0$, to the entire domain of a geometry that is homeomorphic to $\Gamma_0\times [0,1]$ or $\Gamma_0\times S^1$.
While this does not negate the potential value of such topological sweep parameterizations, it does clarify that these swept parameterizations come with no inherent guarantees of bijectivity, as they may in two dimensions.
}

%% file: sections/PaperBody.tex
\section{Introduction}

In computational geometry and computer aided engineering it is often necessary to create a bijective map between domains.
This can be used for data transfer between meshes, for parameterization of meshes, for morphing between shapes, and for mesh generation, among many other applications.
For two-dimensional convex domains, harmonic functions provide guaranteed bijective maps (see \cite{Adamowicz:2020,Weber:2014,Floater:2003}).
More generally, for two-dimensional domains that are not convex, the composition of harmonic maps and their inverse yields a guaranteed bijection between arbitrary two-dimensional domains with disk-like topology \cite{Weber:2014}.
For three-dimensional domains, however, harmonic functions are guaranteed to give bijective maps only when a very restrictive set of conditions are met \cite{Laugesen:1996,DeVerdiere:2003,Floater:2006,Alexa:2025}.

Provided the guarantees available for surfaces and the lack of similar guarantees for volumes, it is natural to wonder if one could leverage the guarantees provided by harmonic functions in lower dimensions to ensure guarantees in some subset of geometries in higher dimensions.
One of the simplest potential generalization leveraging lower dimensional results would be to consider a volume (a 3-manifold with boundary) that is the Cartesian product of a surface and a one-dimensional manifold---either the unit interval $[0,1]$ or the one-sphere, $S^1$---and use guarantees available in the one-dimensional setting to sweep the two-dimensional surface in along this one-manifold using a harmonic map.
Indeed, this method has been proposed in a variety of papers \cite{Gao:2016,Martin:2009}, and some have claimed to provide guaranteed bijective maps using harmonic functions defined in this manner \cite{Xia:2010,Lin:2015}.

The purpose of this paper is to demonstrate through counterexample that such a parameterization of a surface swept along a topological one-manifold does not come with guarantees.
Upon demonstration, we also discuss the lapses in the proofs that make them invalid.

The rest of this paper proceeds as follows:
Section \ref{sec:background} contains background on harmonic functions and Morse theory, which are necessary to understand the proofs.
It also describes the use of harmonic functions to create maps between topological sweeps.
Section \ref{sec:counterexample} gives the computational counterexample to these proofs.
Section \ref{sec:proofs} then discusses in more detail what lapses in the proofs allowed the counterexample, and we follow that up with a brief discussion of the implications of the refutation of these proofs.

\section{Background}\label{sec:background}
\subsection{Manifolds and simplicial complexes}

We discuss both smooth manifolds and their discrete counterparts, piecewise linear simplicial complexes, in the following sections.

A manifold $M$ is a generalization of curves and surfaces to arbitrary dimension.
More precisely, an $n$-dimensional smooth manifold is a topological space such that every point $p\in M$ has a neighborhood that is homeomorphic to an open subset of $\mathbb{R}^n$, and adjacent pairs of these neighborhoods have smooth transition maps that map shared points between them.
Such a smooth manifold locally behaves like Euclidean space, and admits analyses such as calculus.
An $n$-dimensional manifold with boundary has some points, forming the boundary, which instead have a neighborhood homeomorphic to $\mathbb{R}^+\times\mathbb{R}^{n-1}$.
While this precise definition underlies the mathematical theory used later in the paper, we will not apply it directly in this work.
We will abuse terminology and use the term manifold to refer to a smooth manifold with boundary, and denote the boundary of a manifold $M$ as $\partial M$.
We will also denote scalar functions on the manifold as $f:M\rightarrow\mathbb{R}$.

A \textit{simplicial complex} is a piecewise linear object composed of cells called simplices, e.g, points, lines, triangles, and tetrahedra.
An $n$-dimensional simplex is the convex hull of $n+1$ points, excluding cells such as quadrilaterals, hexahedra, and pyramids, and is a closed set.
An important property of a simplicial complex is that it must include all simplices which are boundaries of other cells in the complex.
A pure or homogeneous simplicial $n$-complex denotes a simplicial complex which contains only simplices of dimension $n$ and the lower dimensional boundary cells of these $n$-simplices.
A homogeneous simplicial complex is a type of manifold, though it is not a smooth manifold.
In this work all simplicial complexes are assumed to be homogeneous.
We write a simplicial 3-complex as $\hat{M}=\left\{\mathcal{V},\mathcal{E},\mathcal{F},\mathcal{T}\right\}$, where $\mathcal{V}$, $\mathcal{E}$, $\mathcal{F}$, and $\mathcal{T}$ represent the points, lines, triangles, and tetrahedra, respectively, in the complex.
We also denote any specific dimension $d$ of simplices as $\mathcal{C}^d$.
The star of a cell $a\in \hat{M}$ is defined as $\mathrm{star}(a)=\bigl\{b\in \hat{M}|a\subseteq b\bigr\}$.
We define a dimension-specific star of a cell as $\mathrm{star}^d(a)=\mathrm{star}(a)\cap\mathcal{C}^d$.
The link of a vertex $v_i\in\mathcal{V}$ is defined as
\begin{align*}
	\mathrm{link}(v_i)=\left\{a\in \hat{M}|a\subset c \in \mathrm{star}(v_i) \land a\cap v_i=\varnothing\right\}.
\end{align*}

A discrete scalar function $\hat{f}:\mathcal{V}\rightarrow \mathbb{R}$ is specified by a value $\hat{f}_i$ corresponding to each $v_i\in\mathcal{V}$.
The function is extended to the interior of each $n$-simplex for $n>0$ by linearly interpolating the values at the vertices.
We will denote this piecewise linear (PL) function as $\hat{f}:\hat{M}\rightarrow\mathbb{R}$ as well, and infer which $\hat{f}$ is referred to by the surrounding context.
For a given function $f$, we define the sublevel set as $f^{-1}((-\infty,a])=\{f^{-1}(x)|x\in(-\infty,a]\}$.
We can also define the \textit{lower link} of a vertex $v_i$ with respect to a PL function $\hat{f}$ as $\mathrm{link}_{\hat{f}}^{-}(v_i)=\mathrm{link}(v_i)\cap \hat{f}^{-1}((-\infty,\hat{f}_i])$.

\subsection{Harmonic functions}

A \textit{harmonic function} on $M$ is a function $f: M \rightarrow \mathbb{R}$ that is twice differentiable and satisfies the Laplace equation, $\nabla^2f=0$, where $\nabla^2$ is the Laplace-Beltrami operator for $M$.
Harmonic functions satisfy the maximum principle, which states that a harmonic function on $M$ attains its minimum and maximum on the boundary $\partial M$.

A \textit{discrete harmonic function} is a discrete scalar function that satisfies the discrete Laplace equation,
\begin{align}
	(\mathbf{L}\hat{f})_i=\sum_{e_{ij}\in\mathrm{star}^1(v_i)}w_{ij}(\hat{f}_i-\hat{f}_j)=0 \quad\forall v_i\in\mathcal{V},
\end{align}
where $w_{ij}$ is a weight assigned to the edge $e_{ij}\in\mathcal{E}$ that has as its endpoints $v_i$ and $v_j$.
The choice of weights $w_{ij}$ determine the properties of the discrete Laplace operator, and several choices are available \cite{Alexa:2020,Wardetzky:2007}.
A common choice is the cotangent weights, which creates a discrete Laplace operator that converges to $\nabla^2$ under refinement.
Another option is to choose a set of weights which are non-negative, giving us a discrete maximum principle analogous to the smooth maximum principle \cite{Wardetzky:2007}.
In other words, for non-negative weights $w_{ij}$, the discrete harmonic function $\hat{f}$ attains its maximum and minimum values on the boundary, $\partial \hat{M}$.

\subsection{Morse theory}

\begin{figure}
	\centering
	\begin{subfigure}[t]{0.1\textwidth}
		\caption*{1-saddle:}
		\includegraphics[width=\textwidth]{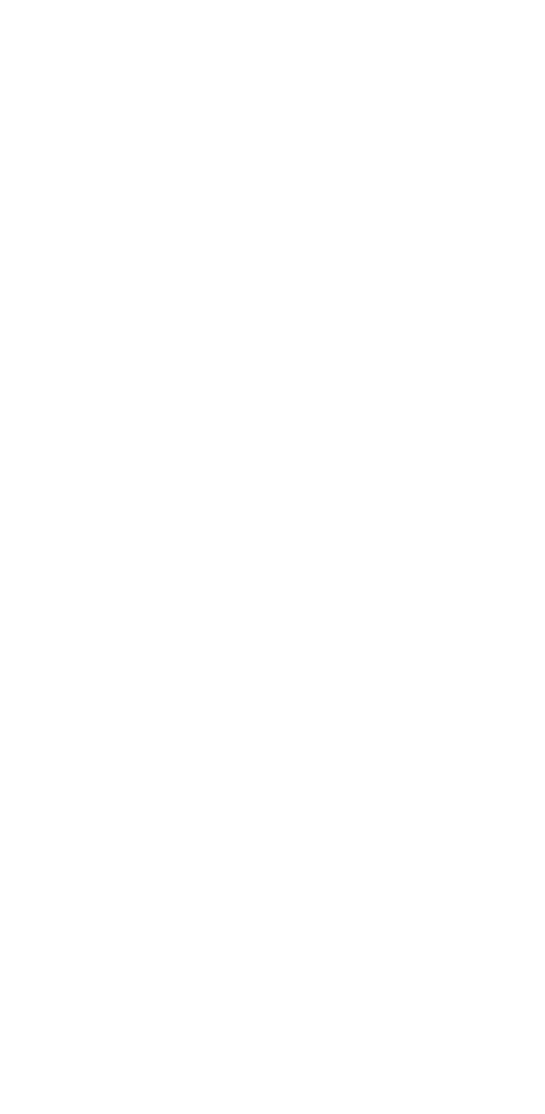}
		\caption*{2-saddle:}
	\end{subfigure}
	\begin{subfigure}[t]{0.2\textwidth}
		\caption*{$a-\varepsilon$}
		\includegraphics[width=\textwidth]{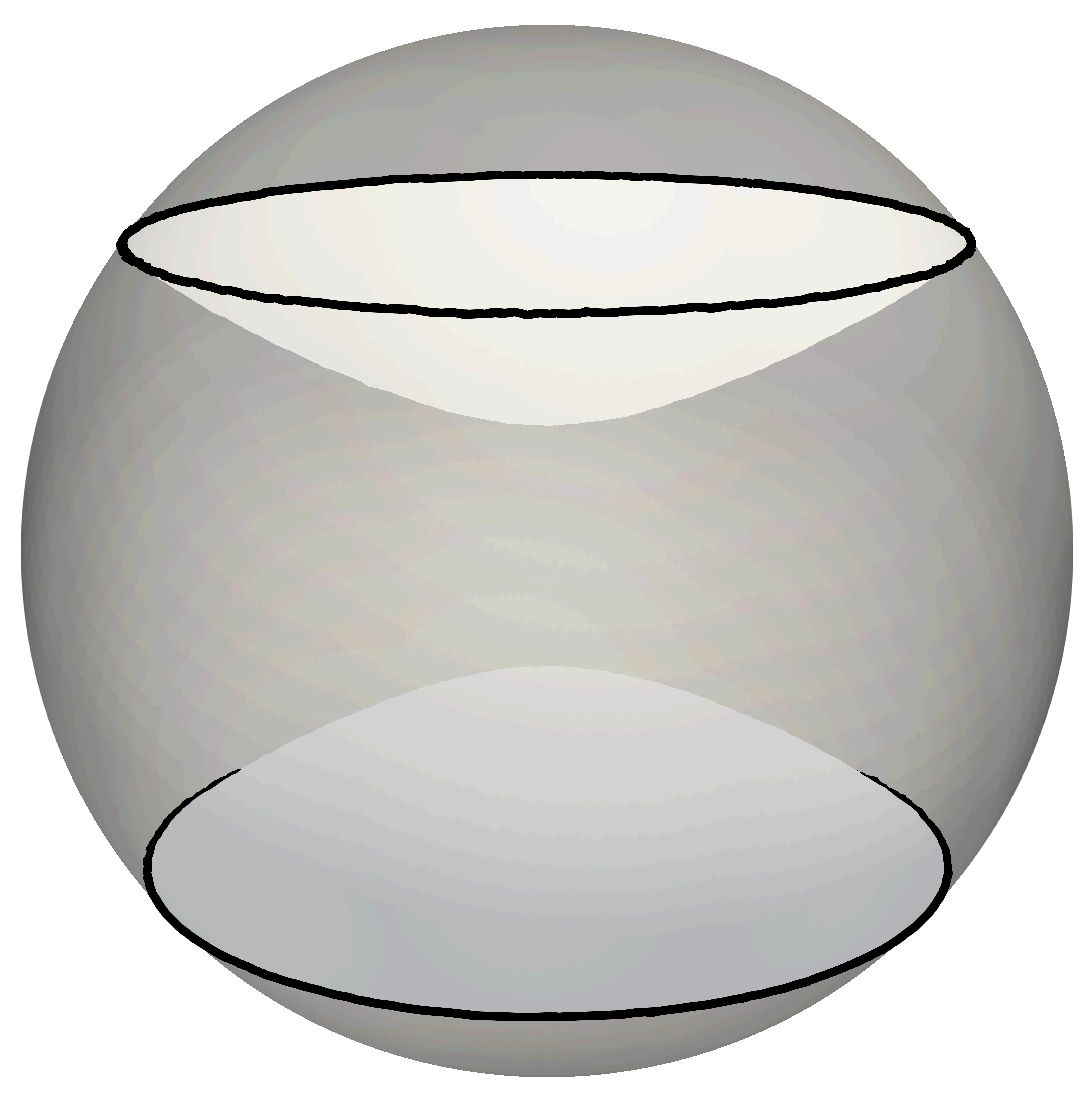}
		\caption*{$a+\varepsilon$}
	\end{subfigure}
	~
	\begin{subfigure}[t]{0.2\textwidth}
		\caption*{$a$}
		\includegraphics[width=\textwidth]{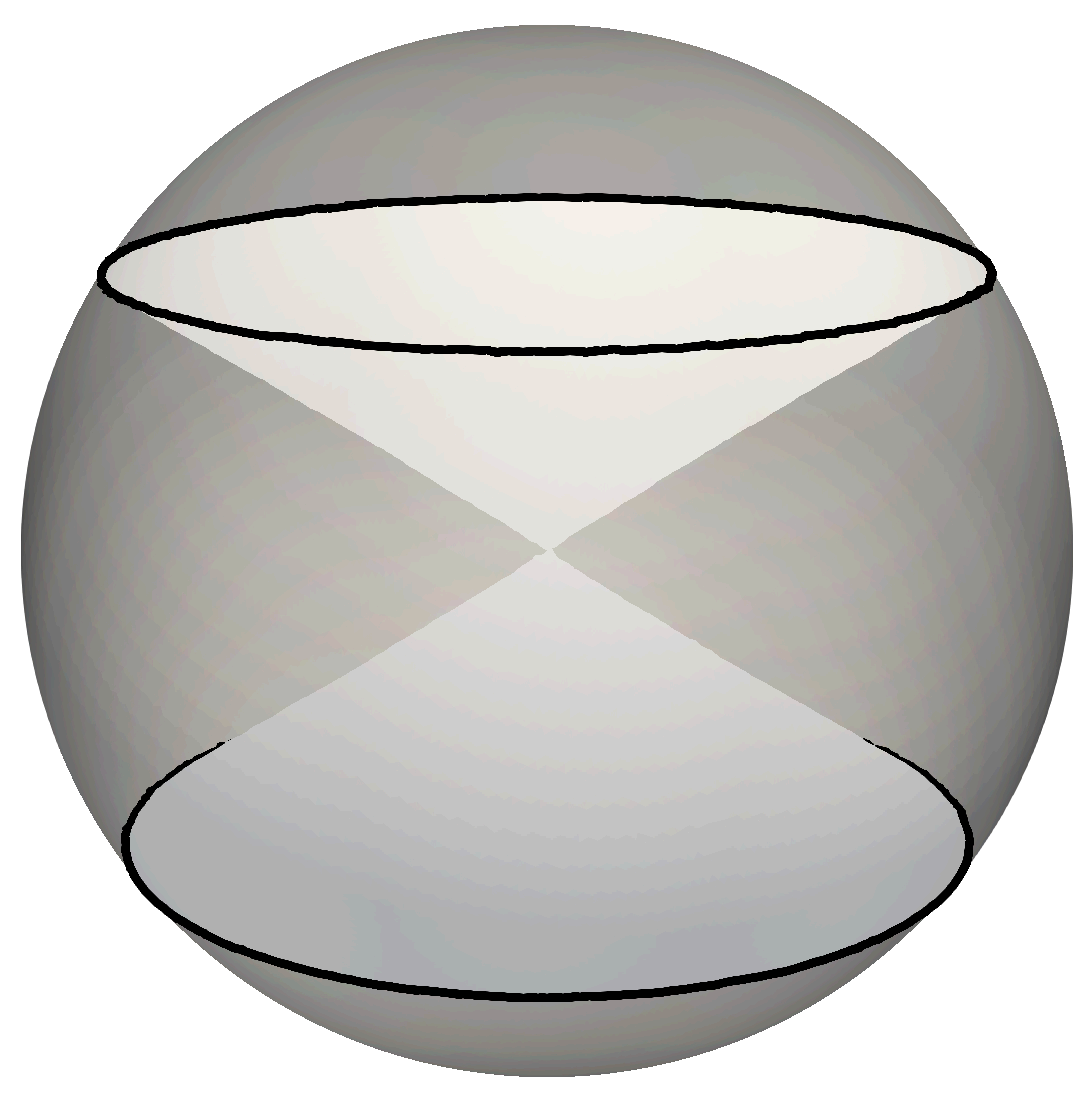}
		\caption*{$a$}
	\end{subfigure}
	~
	\begin{subfigure}[t]{0.2\textwidth}
		\caption*{$a+\varepsilon$}
		\includegraphics[width=\textwidth]{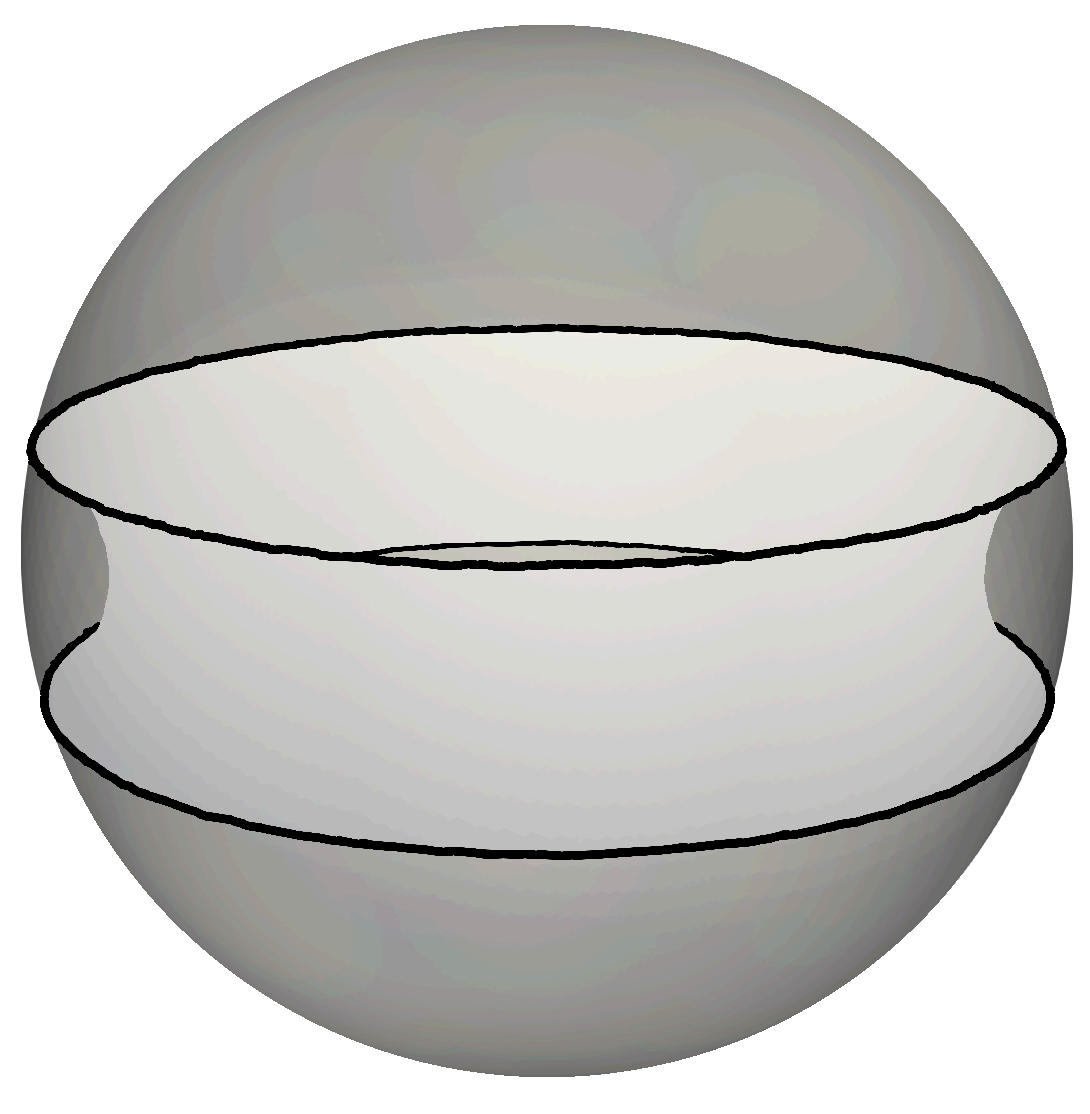}
		\caption*{$a-\varepsilon$}
	\end{subfigure}
	\caption{The local effects of 1-saddles and 2-saddles on the level sets of a function for a critical point with value $a$.}
	\label{fig:criticalpoint}
\end{figure}

A \textit{Morse Function} is a $C^2$ continuous function $f: M \rightarrow \mathbb{R}$ that has only non-degenerate critical points \cite{Hirsch:1994}.
A critical point is any point $\mathbf{p}\in M$ for which $\left\lVert \nabla f(\mathbf{p}) \right\rVert=0$, and a non-degenerate critical point is one at which the Hessian matrix $H(\mathbf{p})=\left[\partial f^2 / \partial x_i \partial x_j\right]$ is nonsingular.
We say that a non-degenerate critical point has index $k$, where $k$ is the number of negative eigenvalues of $H(\mathbf{p})$.
Almost all $C^2$ functions are Morse, and a given function can be made Morse by perturbing it slightly with linear terms in the coordinates of the space \cite{Guillemin:2010}.
Morse functions are very nice in several ways, including that the level sets of these functions only change topology at critical points of $f$ \cite{Knauf:2020}.

There are multiple competing discrete versions of Morse theory, which are equivalent in some ways \cite{Fugacci:2020}.
Because we are particularly working with piecewise linear functions, we will use the PL Morse theory first introduced by Banchoff \cite{Banchoff:1967}.

Critical points in PL Morse theory are always located on vertices of the simplicial complex, and can be defined based on reduced Betti numbers of the lower link of the vertex \cite{Edelsbrunner:2003}.
The reduced Betti numbers $\tilde{\beta}_j$ for $j\in\{-1,0,1,2\}$ of the lower link are the same as the Betti numbers $\beta_j$, which denote the rank of the $j$th homology group of the lower link, except that $\tilde{\beta}_0=\beta_0-1$, and $\tilde{\beta_{-1}}$ is $1$ for an empty lower link and $0$ otherwise.
A vertex $v_i\in\mathcal{V}$ is considered \textit{regular} if $\tilde{\beta}_j=0$ for all $j\in\{-1,...,2\}$, and \textit{critical} otherwise.
We say that a critical vertex is of index $k$ if $\tilde{\beta}_{k-1}\neq0$.
A critical vertex of index 0, 1, 2, or 3 is respectively called a minimum, 1-saddle, 2-saddle, or maximum and is called a multiple saddle or degenerate if $\tilde{\beta}_0 + \tilde{\beta}_1>1$.
We show in Fig.~\ref{fig:criticalpoint} the effects of 1-saddles and 2-saddles on the level sets near the critical point.
Specifically, note that the lower link, which can be taken as the level sets in the first column, has $\tilde{\beta}_0=1$ for the 1-saddle and $\tilde{\beta}_1=1$ for the 2-saddle, as per the definition.
Also of note is that the 1-saddle and 2-saddle have the reverse effect on the level sets; indeed, all 1-saddles of a function $f$ are 2-saddles of its negative, $-f,$ and vice versa.

A PL Morse function is an injective discrete scalar function, $\hat{f}:\hat{M}\rightarrow\mathbb{R}$ with no degenerate PL critical vertices.
Note that the definition of a PL Morse function requires that the function has a different value at each vertex.
In many real-life scenarios, a function that satisfies this criterion can be created from one which does not by vanishingly small perturbations of the values of $\hat{f}$.

Like the smooth setting, PL Morse theory tells us that the topology of level sets of a PL function only changes at PL critical points, i.e., critical vertices \cite{Grunert:2023}.

\subsection{Swept domain harmonic maps}
An important step for generating a volumetric sweep from a surface is to define a harmonic map that acts as the third coordinate of a bijective map.
This is done as follows:
Given a tetrahedral mesh $\hat{M}$ such that $\hat{M}$ is homeomorphic to some $\Gamma_0\times[0,1]$ for $\Gamma_0\subset\partial\hat{M}$, we solve the following discrete Laplace equation for $\hat{f}$:
\begin{align}
	\label{eq:sweep_harmonic}
	\begin{split}
	\sum_{e_{ij}\in\mathrm{star}^1(v_i)}w_{ij}(\hat{f}_i-\hat{f}_j)=0 &\quad\forall v_i \in \mathcal{V}\setminus\Gamma_0\cup\Gamma_1\\
	\hat{f}_i = 0&\quad\forall v_i \in \Gamma_0\\
	\hat{f}_i = 1&\quad\forall v_i \in \Gamma_1,
	\end{split}
\end{align}
where $\Gamma_1\subset\partial\hat{M}$, $\Gamma_0\cap\Gamma_1=\varnothing$, and $\Gamma_0\cong\Gamma_1$.
If $\hat{M}$ is homeomorphic to $\Gamma_0\times S^1$, we do not require that $\Gamma_0\subset\partial\hat{M}$.
Instead, we cut $\hat{M}$ along a subspace of the form $\Gamma_0\times\{p\}$ for some point $p\in S^1$, resulting in a space homeomorphic to $\Gamma_0\times[0,1]$.
We then treat it as described above, with $\Gamma_0$ and $\Gamma_1$ being the two surfaces that resulted from the cut.
We call $\Gamma_0$ the base surface of the sweep, and require that it be parameterized in a 2d sense to create the swept volumetric parameterization.
This volumetric parameterization will have three coordinates, which we will denote as $s$, $t$, and $u$.
The $u$ coordinate of each vertex $v_i$ in the parameterization is taken as $\hat{f}_i$.

The $s$ and $t$ coordinates of the parameterization are found by translating the base surface along the gradient of the harmonic function.
In other words, the $s$ and $t$ coordinates of each vertex are found by tracing the negative of the gradient $\nabla\hat{f}$ from $v_i$ until it reaches the base surface, and the coordinates of that base point in the 2d parameterization become the $s$ and $t$ coordinates in the 3d parameterization.
Another conceptually equivalent method of finding the $s$ and $t$ coordinates is by mapping the coordinates of the base surface to each level set of the harmonic function.
For example, this is done using a simple template applied to a selection of level sets in \cite{Gao:2016}.

It is asserted in \cite{Lin:2015} and \cite{Xia:2010} that there are no critical points in the interior of the domain, and that these traces therefore never intersect, meaning that the resulting sweep parameterization is bijective.
By Morse theory, this is equivalent to stating that the level sets all have equivalent topology.
This requirement on the level sets is necessary for those methods that map to each level set.

\section{Counterexample}\label{sec:counterexample}

\begin{figure}[t]
	\label{fig:example}
	\centering
	\begin{subfigure}[t]{0.4\textwidth}
		\centering
		\includegraphics[width=0.75\textwidth]{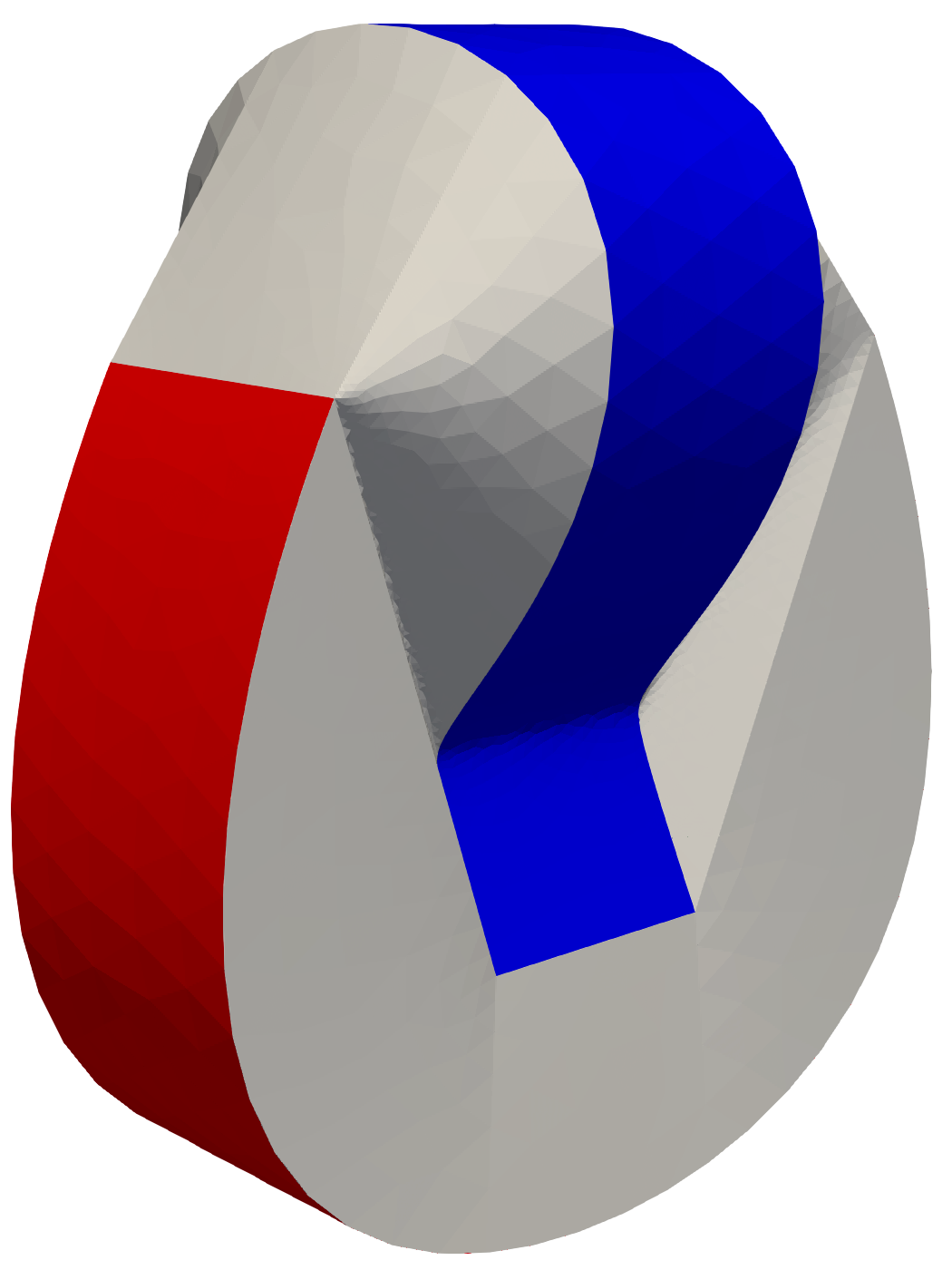}
		\caption{}
		\label{fig:example_geometry}
	\end{subfigure}
	~
	\begin{subfigure}[t]{0.4\textwidth}
		\centering
		\includegraphics[width=0.75\textwidth]{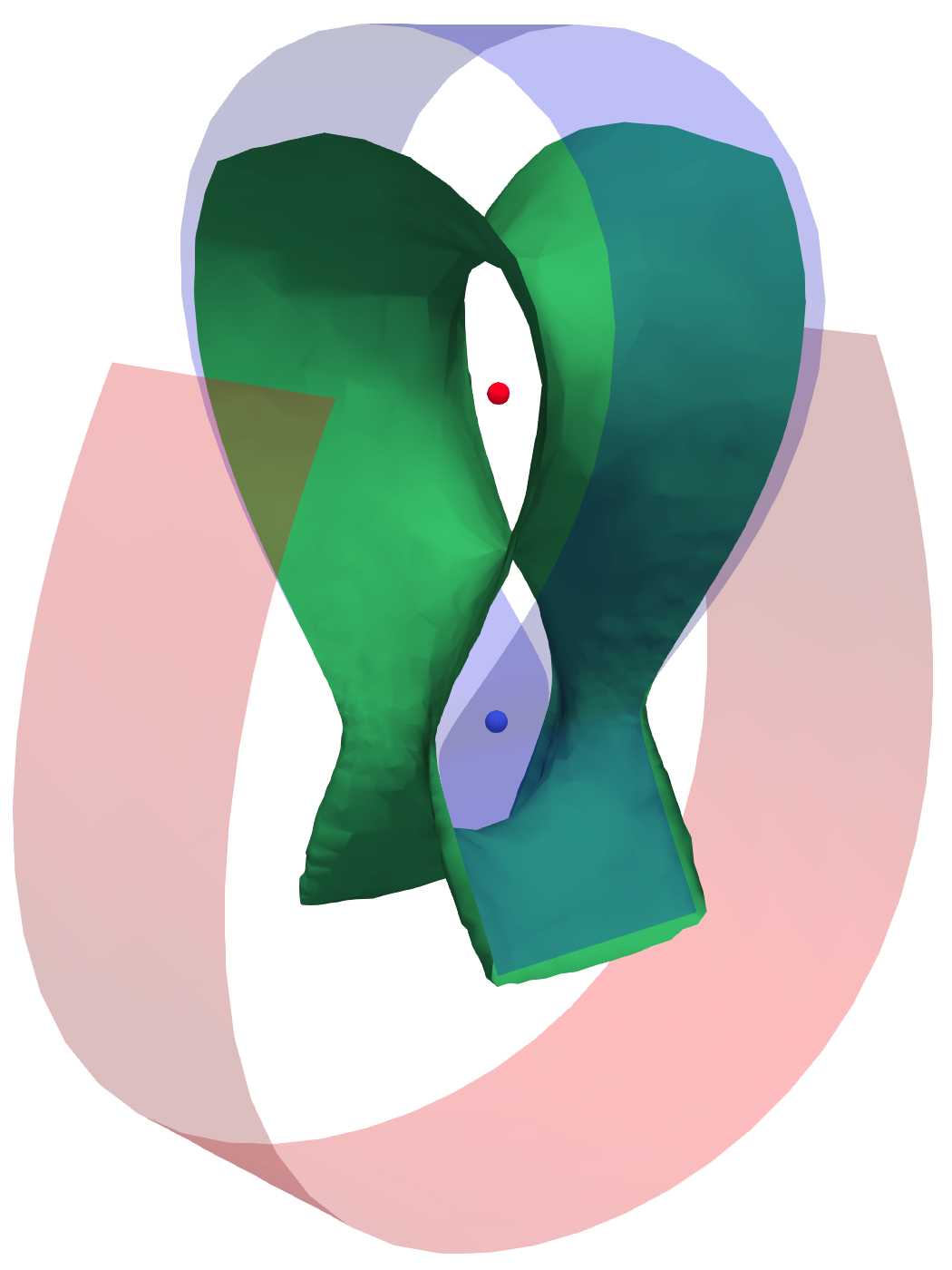}
		\caption{}
		\label{fig:example_levelsets}
	\end{subfigure}
	\caption{
		(a) The geometry of the computational counterexample.
		The geometry is symmetrical in two directions, and not in the third.
		$\Gamma_0$ is indicated in red, and $\Gamma_1$ in blue.
		(b) Level sets of the discrete harmonic function at values of 0.0, 0.85, and 1.0.
		Also shown are the critical points, with an index 1 critical point in red, and an index 2 singularity in blue.
		The level set $\hat{f}^{-1}(0.85)$, shown in green, has nonzero Betti numbers of $\beta_0=1$, $\beta_1=2$, and $\beta_2=0$, while both $\Gamma_0$ and $\Gamma_1$ have $\beta_0=1$, $\beta_1=0$, and $\beta_2=0$.
		}
\end{figure}

We now present a computational counterexample to the proofs mentioned above.
As there is some ambiguity around how to trace the piecewise constant gradient (see, e.g., Fig.~2 in \cite{Ray:2014}), we demonstrate the effects of the critical points in our counterexample using the level sets.

The geometry of the counterexample is shown in Fig.\;\ref{fig:example_geometry} with $\Gamma_0$ marked in red, and $\Gamma_1$ marked in blue.
This shape has symmetry across both the x-z and y-z planes, and was created by first generating the surfaces $\Gamma_0$ and $\Gamma_1$, and then lofting surfaces between their boundaries.
Key attributes of the geometry include that each $\Gamma_i$ encompasses the region between the ends of the other $\Gamma_j$, while its own ends converge slightly.

We compute the sweep harmonic function as given in (\ref{eq:sweep_harmonic}), with the regions $\Gamma_0$ and $\Gamma_1$ as shown in Fig.\;\ref{fig:example_geometry}.
The discrete harmonic weights are the barycentric variation of the dual weights suggested in \cite{Alexa:2020}, which give positive edge weights even on the boundary, thereby causing $\hat{f}$ to satisfy the maximum principle.

In Fig.\;\ref{fig:example_levelsets} we show the level sets $\hat{f}^{-1}(0)$, $\hat{f}^{-1}(0.85)$, and $\hat{f}^{-1}(1)$, along with the critical vertices present in $\hat{f}$.
For this example there are two critical vertices: a 1-saddle in the lower half of the geometry, marked in blue, and a 2-saddle in the upper half, marked in red.
Also, the level sets $\hat{f}^{-1}(0)$ and $\hat{f}^{-1}(1)$ are homeomorphic to a disk, while $\hat{f}^{-1}(0.85)$ is homeomorphic to a torus with a hole in it.

We see that though we satisfy the maximum principle, the geometry is homeomorphic to $\Gamma_0\times[0,1]$, and $\Gamma_0\cong\Gamma_1$, we still have changes of topology in the level sets and critical points in the harmonic function.
This directly contradicts a main point of the proofs in \cite{Lin:2015} and \cite{Xia:2010}.

Note that though we are using a somewhat obscure discrete Laplace operator in order to enforce the maximum principle, this counterexample exhibits the same number and types of critical points with the more common cotangent weights Laplace operator (which does not satisfy the discrete maximum principle).

\section{Previous Proofs}\label{sec:proofs}

Here we describe in more detail the proofs for which we have presented a counterexample, and discuss the lapses in the proofs that allow this counterexample to exist.

First, we discuss the proof in section 4.6 of \cite{Lin:2015}.
This paper maps a tetrahedral mesh of a topological ball to a cube domain using three sweep harmonic functions, $\hat{f}_s$, $\hat{f}_t$, and $\hat{f}_u$, to provide the $s$, $t$, and $u$ coordinates of the parameterization, respectively.
In their proof they assume that the mapping they have constructed in this way is not invertible at a point $\mathbf{p}$, i.e., that the Jacobian matrix $J=(d\hat{f}_s,d\hat{f}_t,d\hat{f}_u)^T$ is singular.
This assumption leads to the conclusion that $\mathbf{p}$ is a critical point of a harmonic function $\hat{f}$ constructed as a linear combination of the $\hat{f}_i,\;i\in\{s,t,u\}$.
The conclusion in the paper is that this is a contradiction, because of the maximum principle.
This is simply an error of neglecting the possibility of the critical point being a 1-saddle or a 2-saddle.

The proof in section IV-E of \cite{Xia:2010} requires a more nuanced discussion.
Here the authors attempt to prove that lines tracing the gradient of the sweep harmonic function from the base surface cannot merge on the interior of the domain.
This is an essential point in their proof that the sweep harmonic map is bijective.
They mention that for these lines to merge, there must be a critical point at the merge, which is true in the continuous setting.
Since maxima and minima are excluded by the maximum principle, the critical point must be a 1-saddles or a 2-saddle.
Recall the result from Morse theory that critical points change the topology of the level sets.
The proof notes that $\hat{f}^{-1}(0)=\Gamma_0\cong\Gamma_1=\hat{f}^{-1}(1)$ and that \textit{the topology change of the level sets from critical points cannot be canceled out}.
Therefore, there must be no critical points on the interior.

We specifically look at the statement that the topology change of the level sets from critical points cannot be canceled out.
From Lemma 2.4, and specifically the following discussion in Remark 2.6, of \cite{Knauf:2020} we see that a 1-saddle causes a change in the Betti numbers of the level set either of $\Delta\beta_1=2$ and the remaining $\Delta\beta_k=0$, or $\Delta\beta_0=\Delta\beta_2=-1$ and the remaining $\Delta\beta_k=0$.
On the other hand, a 2-saddle causes a change in the Betti numbers of the level set either of $\Delta\beta_1=-2$ and the remaining $\Delta\beta_k=0$, or $\Delta\beta_0=\Delta\beta_2=1$ and the remaining $\Delta\beta_k=0$.
While these are not sure to cancel each other out, they can, if the cases match.
We see in the counterexample above that the 1-saddle increases $\beta_1$ by 2, and the 2-saddle decreases $\beta_1$ by 2, so the topology change of the level sets is canceled out in this case.

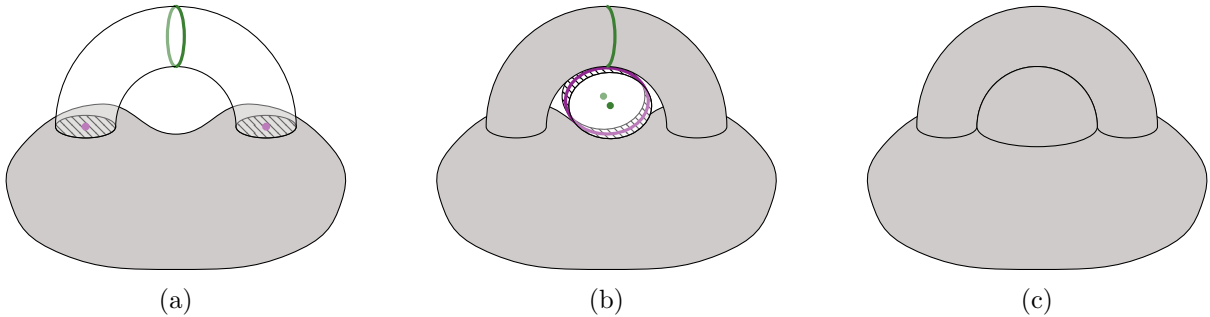
\begin{figure}
	\centering
	\input{figures/handle_cancellation.tex}
	\caption{Cancellation of a 1-handle and a 2-handle attached to a 3-manifold.
			Attaching regions are filled with hash lines, The attaching sphere of each handle is shown in purple, and the belt sphere of each handle is shown in green.
			(a) Attaching a 1-handle to the manifold.
			(b) Attaching a 2-handle to the manifold and one-handle. Notice that the attaching sphere of the 2-handle intersects the belt sphere of the 1-handle at a single point.
			(c) The result of the handle attachment is homeomorphic to the original manifold.
	}\label{fig:handle_cancellation}
\end{figure}

\begin{figure}
	\centering
	\begin{subfigure}[t]{0.3\textwidth}
		\includegraphics[width=\textwidth]{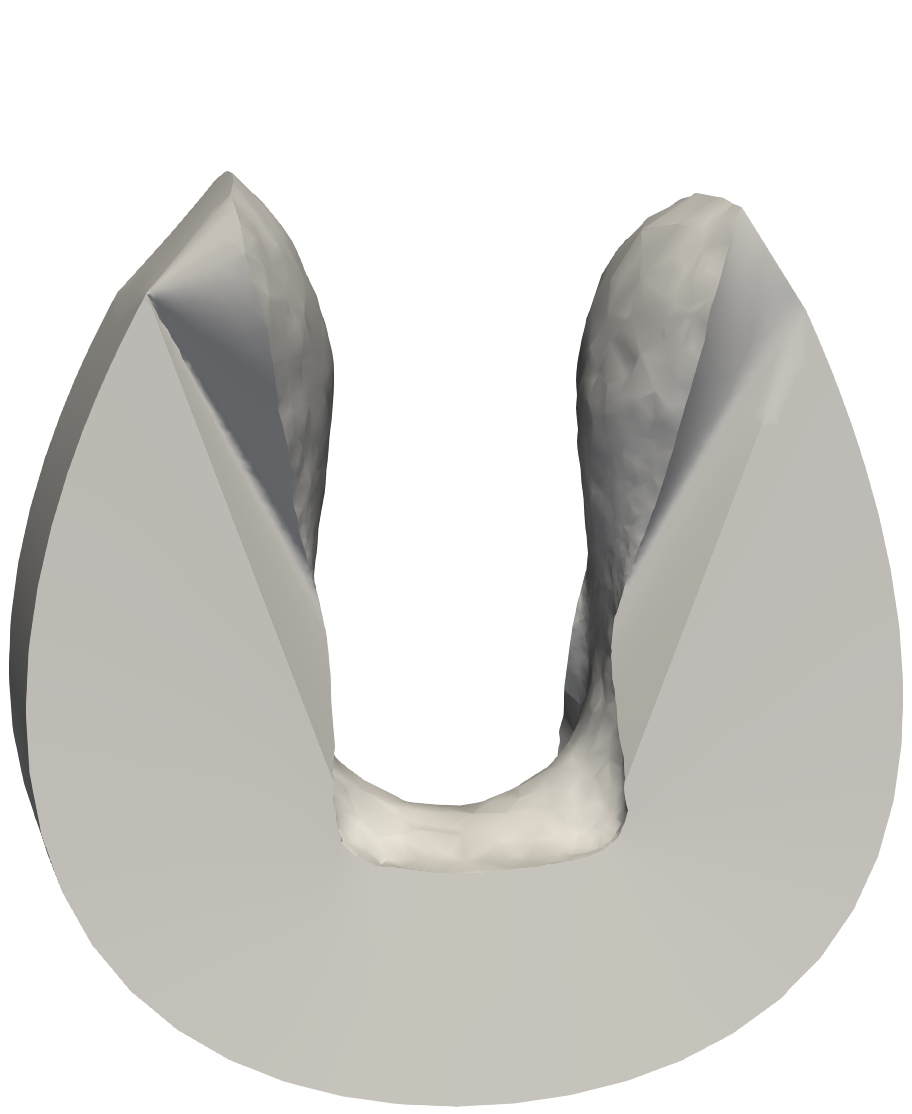}
		\caption{}
	\end{subfigure}
	~
	\begin{subfigure}[t]{0.3\textwidth}
		\includegraphics[width=\textwidth]{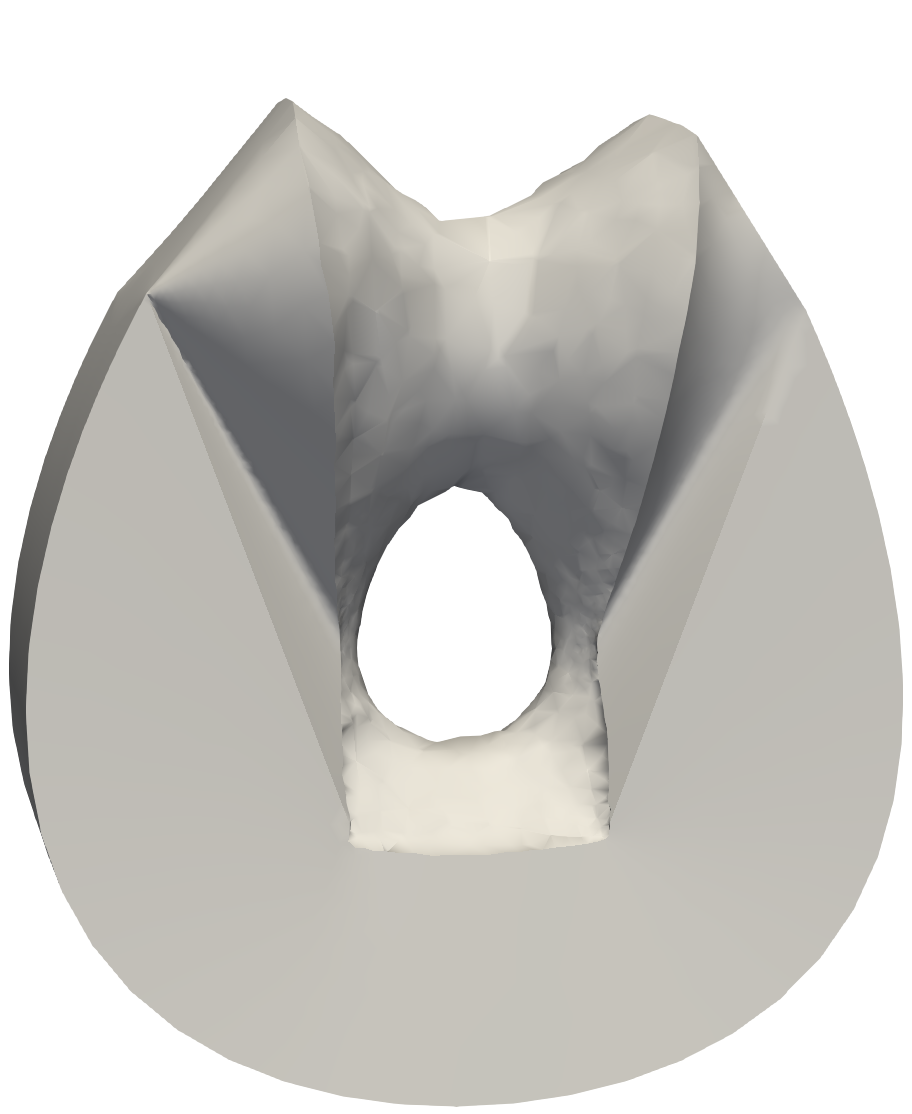}
		\caption{}
	\end{subfigure}
	~
	\begin{subfigure}[t]{0.3\textwidth}
		\includegraphics[width=\textwidth]{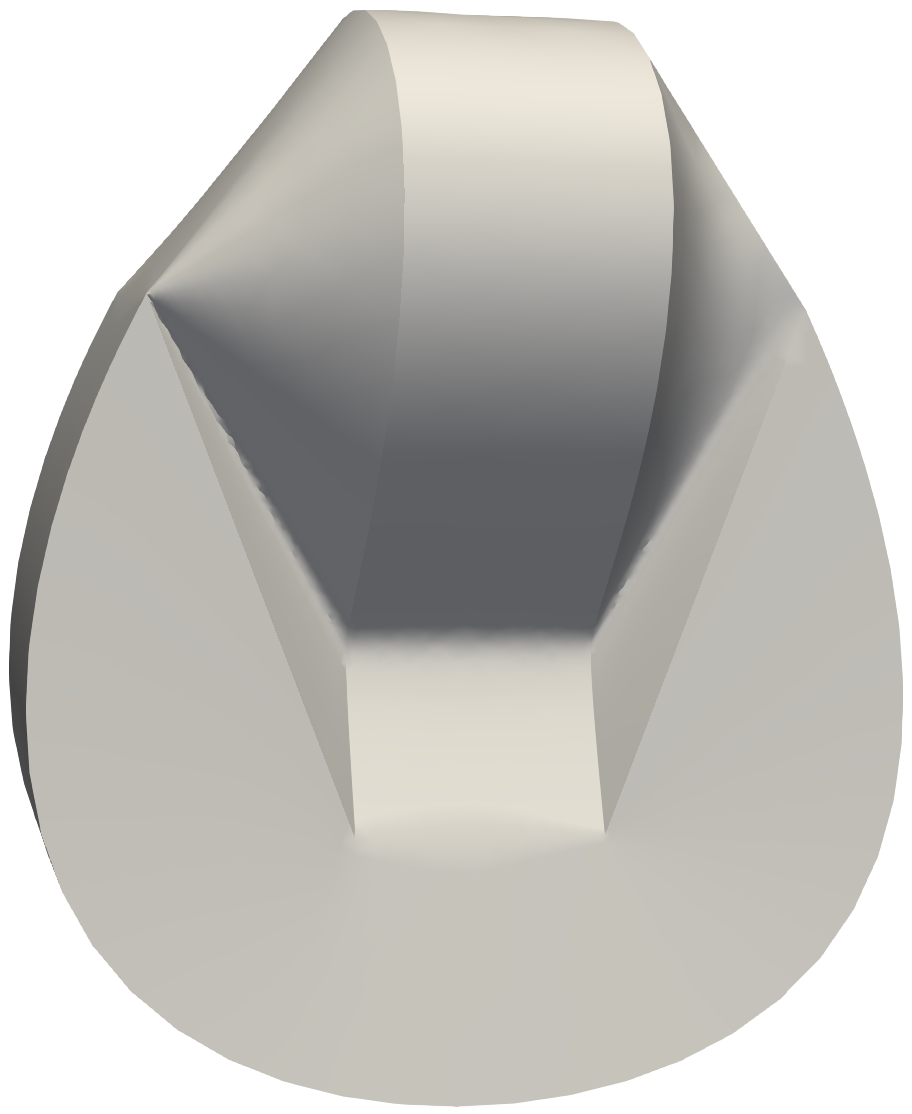}
		\caption{}
	\end{subfigure}
	\caption{Sublevel sets of the discrete harmonic function on the counterexample.
	Compare to Fig.~\ref{fig:handle_cancellation}.
	(a) $\hat{f}^{-1}([0,0.7])$.
	(b) $\hat{f}^{-1}([0,0.85])$.
	(c) $\hat{f}^{-1}([0,1])=\hat{M}$.
	}\label{fig:sublevel_sets}
\end{figure}

For the sweep case specifically, the requirement that all level sets are homeomorphic to $\Gamma_0$ is equivalent to requiring that each sublevel set $\hat{f}^{-1}([0,a])$ for $0<a<=1$ is homeomorphic to $\Gamma_0\times[0,1]$.
This requirement allows for an alternate description of how the critical points cancel each other based on handlebody theory.
Specifically, if $\hat{f}^{-1}(a)$ contains one $k$-index critical point, the sublevel set $\hat{f}^{-1}([0,a+\varepsilon])$ is diffeomorphic to $\hat{f}^{-1}([0,a-\varepsilon])$ with a $k$-handle attached \cite{Knauf:2020}.
Fig.~\ref{fig:handle_cancellation} shows a diagram of this process, and Fig.~\ref{fig:sublevel_sets} shows the analogous process for the sublevel sets of the counterexample.
The handle cancellation theorem from handlebody theory tells us that the attachment of a $k$-handle followed by a $(k+1)$-handle to a space $M$ is homeomorphic to $M$ if the handles intersect in a certain way (for details, see \cite{Rourke:1972}).
This condition is equivalent to the green curve in the 1-handle in Fig.~\ref{fig:handle_cancellation} intersecting the purple curve on the 2-handle at a single point.
Note that a 1-handle can also cancel with a 0-handle under this theorem, and a 2-handle can also cancel with a 3-handle.
These handles are attached by a minimum and a maximum, respectively, however, and so the maximum principle precludes this kind of cancellation as long as a discrete operator that satisfies the maximum principle is chosen.

This sublevel set perspective allows us to contrast with the 2-dimensional case.
In 2d, there is only one type of saddle, a 1-saddle.
This saddle adds a 1-handle, but there is no other critical point on the interior to cancel that handle because of the maximum principle, so for the sweep compatible geometry there can be no 1-saddles and therefore no critical points on the interior.

\section{Conclusion}
We have shown here a counterexample to the proofs that have been presented for bijective sweep maps in 3d.
Specifically, we have shown that in a geometry that is homeomorphic to $\Gamma_0\times[0,1]$, a sweep harmonic function that obeys the discrete maximum principle can have critical points in the interior of the domain.

While this refutes any guarantees of bijectivity when using the sweep harmonic function as one coordinate of a map, it does not negate the value of these maps in cases where they do work.

%% file: figures/handle_cancellation.tex
\begin{subfigure}[t]{0.3\textwidth}
  \centering
\begin{tikzpicture}[scale=1.0]
    \tikzmath{
    \rh = 0.4;
    \ri = 0.8;
    \rm = \ri + \rh;
    \ro = \rm + \rh;
    \handleHeight=1.2;
    \rhy=0.15;
    \riy=0.8;
    \roy=\riy+2*\rh; } 
    \draw[fill=lightgray!50] plot[smooth cycle, tension=0.8] coordinates {
        (-2,1.0) (-1,1.5) (0,1.1) (1,1.5) (2,1.0) (2.2,0.2) (1.5,-0.5) (0,-0.7) (-1.5,-0.5) (-2.2,0.2)
    };
    
    \draw[pattern=north west lines] (\rm,\handleHeight) ellipse ({\rh} and {\rhy});
    \draw[pattern=north west lines] (-\rm,\handleHeight) ellipse ({\rh} and {\rhy});
    \draw[OliveGreen, very thick](0.0, \handleHeight+\roy)
      arc [x radius=0.75*\rhy, y radius=\rh,start angle=90, end angle=270];

    \node[draw,Plum,very thick,circle, fill=Plum, inner sep=0pt, minimum width=1.8pt] (mycenter) at (\rm,\handleHeight) {};
    \node[draw,Plum,very thick,circle, fill=Plum, inner sep=0pt, minimum width=1.8pt] (mycenter) at (-\rm,\handleHeight) {};

    \draw[fill=white, fill opacity=0.4] (-\ro,\handleHeight)
      arc [x radius=\ro, y radius=\roy, start angle=180, end angle=0]
      arc [x radius=\rh, y radius=\rhy, start angle=0, end angle=-180]
      arc [x radius=\ri, y radius=\riy, start angle=0, end angle=180]
      arc [x radius=\rh, y radius=\rhy, start angle=0, end angle=-180];

    \draw[OliveGreen, very thick] (0.0,\handleHeight+\riy)
      arc [x radius=0.75*\rhy, y radius=\rh,start angle=-90, end angle=90];
\end{tikzpicture}
\caption{}
\end{subfigure}
~
\begin{subfigure}[t]{0.3\textwidth}
  \centering
\begin{tikzpicture}[scale=1.0]
    \tikzmath{
    \rh = 0.4;
    \ri = 0.8;
    \rm = \ri + \rh;
    \ro = \rm + \rh;
    \handleHeight=1.2;
    \rhy=0.15;
    \riy=0.8;
    \roy=\riy+2*\rh;
    \crx=1;
    \cry=0.8;
    \cdz=0.28;
    \cangle=-150;
    \cphi=atan(-\cry/\crx/tan(\cangle));
    \startx=\crx*cos(\cangle);
    \starty=\cry*sin(\cangle);
    } 
    \draw[fill=lightgray!50] plot[smooth cycle, tension=0.8] coordinates {
        (-2,1.0) (-1,1.5) (0,1.1) (1,1.5) (2,1.0) (2.2,0.2) (1.5,-0.5) (0,-0.7) (-1.5,-0.5) (-2.2,0.2)
    };

    \begin{scope}[shift={(0.05 cm,1.48 cm)},scale=0.55]
    \draw[pattern=north west lines] (\startx,\starty)
      arc [x radius=\crx, y radius=\cry,start angle=\cangle, end angle=-180+\cangle]
      --++ (\cphi+180:\cdz)
      arc [x radius=\crx, y radius=\cry,start angle=-180+\cangle, end angle=\cangle]
      -- cycle;
    \draw[fill=white, fill opacity=0.8,draw opacity=0.0] (\startx,\starty)
      arc [x radius=\crx, y radius=\cry,start angle=\cangle, end angle=180+\cangle]
      --++ (\cphi+180:\cdz)
      arc [x radius=\crx, y radius=\cry,start angle=180+\cangle, end angle=\cangle]
      -- cycle;
    \draw[pattern=north west lines] (\startx,\starty)
      arc [x radius=\crx, y radius=\cry,start angle=\cangle, end angle=180+\cangle]
      --++ (\cphi+180:\cdz)
      arc [x radius=\crx, y radius=\cry,start angle=180+\cangle, end angle=\cangle]
      -- cycle;
    \draw[Plum,very thick](\startx,\starty) ++(\cphi+180:\cdz/2) arc [x radius=\crx, y radius=\cry,start angle=\cangle, end angle=360+\cangle];
    \node[draw,very thick,OliveGreen,circle, fill=OliveGreen, inner sep=0pt, minimum width=1.5pt] (mycenter) at (\cphi+180:\cdz) {};
    \draw[fill=white, fill opacity=0.4](\startx,\starty) arc [x radius=\crx, y radius=\cry,start angle=\cangle, end angle=360+\cangle];
    \node[draw,very thick,OliveGreen,circle, fill=OliveGreen, inner sep=0pt, minimum width=1.5pt] (mycenter) at (0,0) {};
    \end{scope}

    \draw[fill=lightgray!50, fill opacity=1.0] (-\ro,\handleHeight)
      arc [x radius=\ro, y radius=\roy, start angle=180, end angle=0]
      arc [x radius=\rh, y radius=\rhy, start angle=0, end angle=-180]
      arc [x radius=\ri, y radius=\riy, start angle=0, end angle=180]
      arc [x radius=\rh, y radius=\rhy, start angle=0, end angle=-180];

    \draw[OliveGreen,very thick] (0.0,\handleHeight+\riy)
      arc [x radius=0.75*\rhy, y radius=\rh,start angle=-90, end angle=90];
\end{tikzpicture}
\caption{}
\end{subfigure}
~
\begin{subfigure}[t]{0.3\textwidth}
  \centering
\begin{tikzpicture}[scale=1.0]
    \tikzmath{
    \rh = 0.4;
    \ri = 0.8;
    \rm = \ri + \rh;
    \ro = \rm + \rh;
    \handleHeight=1.2;
    \rhy=0.15;
    \riy=0.8;
    \roy=\riy+2*\rh;
    \crx=1;
    \cry=0.8;
    \cdz=0.28;
    \cangle=-150;
    \cphi=atan(-\cry/\crx/tan(\cangle));
    \startx=\crx*cos(\cangle);
    \starty=\cry*sin(\cangle);
    } 
    \draw[fill=lightgray!50] plot[smooth cycle, tension=0.8] coordinates {
        (-2,1.0) (-1,1.5) (0,1.1) (1,1.5) (2,1.0) (2.2,0.2) (1.5,-0.5) (0,-0.7) (-1.5,-0.5) (-2.2,0.2)
    };

    \draw[fill=lightgray!50, fill opacity=1.0] (-\ro,\handleHeight)
      arc [x radius=\ro, y radius=\roy, start angle=180, end angle=0]
      arc [x radius=\rh, y radius=\rhy, start angle=0, end angle=-180]
      arc [x radius=\ri, y radius=\riy, start angle=0, end angle=180]
      arc [x radius=\rh, y radius=\rhy, start angle=0, end angle=-180];

    \draw[fill=lightgray!50] (\ri,\handleHeight)
      arc [x radius=\ri, y radius=\riy, start angle=0, end angle=180]
      arc [x radius=\ri, y radius=\riy/3, start angle=-180, end angle=0];
\end{tikzpicture}
\caption{}
\end{subfigure}